\documentclass[psfig,epic,eepic,pstricks,pstricks,pst-plot,
               epsf,pst-node,12pt]{book}
\makeatletter
\def\nothing#1{}
\newdimen\earraycolsep
\def\eqnarray{\let\@currentlabel\theequation
\global\@eqnswtrue\m@th
\global\@eqcnt\z@\tabskip\@centering\let \\\@eqncr
$$\halign to\displaywidth\bgroup\@eqnsel\hskip\@centering
$\displaystyle\tabskip\z@{##}$&\global\@eqcnt\@ne
\hskip 2\earraycolsep \hfil$\displaystyle{##}$\hfil
&\global\@eqcnt\tw@ \hskip 2\earraycolsep $\displaystyle\tabskip\z@{##}$\hfil
\tabskip\@centering&\llap{##}\tabskip\z@\cr}
\renewcommand{\theequation}{\arabic{equation}}
\renewcommand{\thetable}{\arabic{table}}
\renewcommand{\thefigure}{\arabic{figure}}
\def\title{\chapter}
\renewcommand\chapter{\ifodd\c@page\clearpage\else\cleardoublepage\fi
                    \global\@topnum\z@
                    \@afterindenttrue
                    \secdef\@chapter\@schapter}
\def\@makechapterhead#1{%
  \vspace*{120\p@}%
  {\parindent \z@ \raggedright \reset@font
    \bfseries #1\par
    \nobreak
    \vskip 36\p@
  }}
\def\author#1{{\pretolerance=10000 \raggedright \advance \leftskip by 1in
\noindent #1 \vskip 1pc}}
\def\affiliation#1{{\advance\leftskip by 1in \noindent #1 \vskip -1pc}}
\def\refnote#1{{$^{\hbox{\scriptsize #1}}$}}

\renewcommand\section{\@startsection{section}{1}{\z@}{2pc \@plus 1ex minus
    .2ex}{1pc \@plus .2ex}{\reset@font\normalsize\bfseries}}
\renewcommand\subsection{\@startsection{subsection}{2}{\z@}{1pc \@plus 1ex
    minus.2ex}{1pc \@plus .2ex}{\reset@font\normalsize\bfseries}}
\renewcommand\subsubsection{\@startsection{subsubsection}{3}{\parindent}
        {1pc \@plus 1ex minus.2ex}{-0.5em}{\reset@font\normalsize\bfseries}}

\def\AmS{{\protect\the\textfont2
A\kern-.1667em\lower.5ex\hbox{M}\kern-.125emS}}

\def\p@LaTeX{{\family{times}\series{m}\shape{n}\selectfont
L\kern-.36em\raise.3ex\hbox{\scriptsize A}\kern-.15em
T\kern-.1667em\lower.7ex\hbox{E}\kern-.125emX}}

\newlength{\colwidth}

\def\@oddhead{\hfil}
\def\@evenhead{\hfil}
\def\@oddfoot{{\bfseries\hfil\thepage}}
\def\@evenfoot{{\bfseries\thepage\hfil}}
\def\fnum@figure{\footnotesize\raggedright{\bfseries \figurename~\thefigure.}}
\def\fnum@table{\normalsize\raggedright{\bfseries \tablename~\thetable.}}
\long\def\@makecaption#1#2{\vskip 10\p@ {#1 #2\par}}
\long\def\@makefntext#1{\setbox0=\hbox{$\m@th^{\@thefnmark}$}\noindent\hang
indent=\wd0 \box0 #1}
\newbox\@atbox
\long\def\atable#1#2#3{\begin{table}[tbp]\centering\footnotesize
\setbox\@atbox\hbox{#2}
\parbox{\wd\@atbox}{\caption{#1}}\par\smallskip
#2
\par\smallskip\parbox{\wd\@atbox}{\raggedright #3}
\end{table}}
\def\@bibitem{\noindent \hangindent=2pc \hangafter=1}
\def\thebibliography{%
\section*{REFERENCES}%
\bgroup\footnotesize
\def\newblock{\hskip .11em plus.33em minus.07em}%
\let\bibitem\@bibitem}
\def\endthebibliography{\par\egroup}
\def\@nbibitem#1{\noindent \hangindent=2pc \hangafter=1
\refstepcounter{enumi}\hbox to 2pc{\arabic{enumi}.\hfil}%
\immediate\write\@auxout{\string\bibcite{#1}{\arabic{enumi}}}}
\def\numbibliography{%
\section*{REFERENCES}%
\bgroup\footnotesize
\setcounter{enumi}{0}%
\def\newblock{\hskip .11em plus.33em minus.07em}%
\let\bibitem\@nbibitem}
\def\endnumbibliography{\par\egroup}
\makeatother

\begin{document}
\chapter{Density-Functional Theory of Surface Diffusion and \\
Epitaxial Growth of Metals}
\author{C. Ratsch, P. Ruggerone, and M. Scheffler}
\affiliation{Fritz-Haber-Institut der Max-Planck-Gesellschaft \\ Faradayweg 4-6, D-14195
Berlin-Dahlem, Germany}
\vspace{.7cm}
This paper gives a summary of basic concepts of 
density-functional theory (DFT) and its use in
state-of-the-art computations of complex processes in condensed matter
physics and  materials science. 
In particular we discuss how microscopic growth parameters can be determined
by DFT and how on this basis macroscopic phenomena can be described.
To reach the time and length
scales of realistic growth conditions, DFT results are complemented
by kinetic Monte Carlo simulations. 
\vspace{1.cm}

\section{INTRODUCTION}

The microscopic processes governing epitaxial growth typically 
are rather complex since they involve the making and breaking of chemical bonds as 
well as the dynamics of the nuclei. With recent progress in the developments
of new methods and techniques and the availability of faster computers,
density-functional theory (DFT) calculations have evolved into a powerful 
tool to study growth phenomena (as well as other complex processes in
condensed matter physics, materials science, and chemistry). 
The rate of a microscopic process $j$ that occurs during growth, such as
diffusion, 
usually has the form $\Gamma^{(j)}=\Gamma_0^{(j)} \exp(-E_{\rm d}^{(j)}/k_{\rm B}T)$,
where $\Gamma_0^{(j)}$ is the effective attempt frequency, $T$ the temperature, $k_{\rm B}$ the
Boltzmann constant, and $E_{\rm d}^{(j)}$ is the energy barrier that needs to be
overcome for the event $j$ to take place.
This equation reflects the idea that an adatom experiences many
stable and metastable sites at the surface, and that the diffusive motion
brings it from one minimum to an adjacent
one on the free energy surface in the space of the reaction coordinates. The
effective attempt frequency $\Gamma_0^{(j)}$ contains the terms due
to the adatom and substrate vibrations 
(see. Ref.~1 for more details). 

It is at the heart of theoretical studies of surface diffusion
and growth phenomena to calculate the ground-state total
energy of the adsorbate system for a dense mesh of adatom positions. This 
yields the so-called potential-energy surface (PES) 
which is the potential energy experienced by the diffusing adatom,
\begin{equation}
E^{\rm PES}(X_{\rm ad}, Y_{\rm ad}) = \min_{Z_{\rm ad},\{{\bf R}_I\}}
E^{\rm tot} (X_{\rm ad}, Y_{\rm ad}, Z_{\rm ad},\{{\bf R}_I\}) \quad,
\label{PES}
\end{equation}
where $E^{\rm tot} (X_{\rm ad}, Y_{\rm ad}, Z_{\rm ad},\{{\bf R}_I\})$
is the ground-state energy of the many-electron system 
(also referred as the total energy)
at the atomic configuration $(X_{\rm ad}, Y_{\rm ad}, Z_{\rm ad},\{{\bf R}_I\})$.
According to Eq.~(\ref{PES}) the PES is the minimum of the 
total energy with respect to 
the $z$-coordinate of the adatom and all coordinates of the substrate atoms
$\{{\bf R}_I\}$. 
Assuming that vibrational effects can be neglected, the minima of the
PES represent stable and metastable sites of the adatom.
Note, this PES refers to slow motion of nuclei and
assumes that for any atomic configuration the
electrons are in their respective  ground state. Thus, it is assumed that
the dynamics of the electrons and of the nuclei are decoupled. This
is the Born-Oppenheimer approximation which for not too high temperatures
is usually well justified.

Now consider all possible paths $j$ to get from one stable or metastable
adsorption site, ${\bf R}_{\rm ad}$, to an adjacent one, ${\bf R}_{\rm ad}{\bf '}$.
The energy difference $E_{\rm d}^j$ between the energy at the saddle point
along $j$ and the energy of the start geometry is the barrier 
for this particular path. The diffusion barrier then is the minimum value of all
$E_{\rm d}^j$ of all possible paths which connect ${\bf R}_{\rm ad}$ and ${\bf R}_{\rm ad}{\bf '}$,
and the lowest energy saddle point is called the transition state.
We note this definition strictly applies only to cases
where the vibrational energy is negligible, which is typically
justified when the diffusion barrier is not too small, and the
temperature is not too high. We also note that for
non zero temperatures, i.e., when vibrations are thermally excited,
vibrational entropy needs to be considered, which will enter
the attempt frequency of the hopping rate.
Although often only the path with
the most favorable energy barrier is important, it may happen that
several paths exist with comparable barriers or that 
the PES consists of more than one sheet
(e.g., Ref.~2).
Then the {\em effective} barrier measured in an experiment (or a molecular
dynamics simulation) represents a proper average  over  all possible pathways.
The above description obviously applies to simple jumps of an
adatom (i.e., diffusion by hopping), but it also holds  for the 
diffusion by atomic exchange where the diffusing adatom displaces a substrate
atom.\refnote{3-7}

Aiming at a calculation and understanding of the PES of a diffusing atom,
it is obvious that the interplay between the breaking and making
of chemical bonds and the atomic relaxations needs to be accounted for by
an accurate, quantum-mechanical 
description of the many-electron
system. This can be achieved by modern density-functional
theory calculations that combine electronic self-consistency
and efficient geometry optimization (see for
example Ref. 8-10 and references therein).

In the following two sections we describe briefly the basic concepts of
density-functional theory (DFT) and of its application to surface problems. We
then discuss some examples where DFT has been used to calculate growth
parameters and compare the results with experiments. Finally, 
an example is given that shows how DFT calculations can be used in 
combination with kinetic Monte Carlo simulations to describe and analyze
epitaxial growth.

\section{DENSITY-FUNCTIONAL THEORY: BASIC CONCEPTS}

The total energy of an $N$-electron, poly-atomic system
is given by the expectation value of the many-particle Hamiltonian
using the many-body wave-function of the electronic ground state.
For a solid or a surface the calculation of such an expectation value is impossible
when using a wave-function approach. However, as has been shown by Hohenberg
and Kohn,\refnote{11}
the ground-state total energy can also be obtained without explicit
knowledge of the many-electron wave-function, but by minimizing
an energy functional $E[n]$.
This is the essence of density-functional theory (DFT),
which is primarily (though in principle not exclusively)
a theory of the electronic ground state, couched in
terms of the electron density $n({\bf r})$  instead of the many-electron
wave-function $\Psi(\{{\bf r}_i\})$.

The important theorem of Hohenberg and Kohn\refnote{11}
(see also Levy\refnote{12}) tells: The specification of a ground state
density $n({\bf r})$ determines the corresponding external potential
$v^{\rm ext}({\bf r})$ $uniquely$ (to within an additive
constant), 
\begin{equation}
n({\bf r}) \to v^{\rm ext}({\bf r})\quad .
\label{defrel}
\end{equation}
The external potential $v^{\rm ext}({\bf r})$ 
is typically (and definitely for our purpose here) the Coulomb
potential due to the nuclei.  While the other direction 
[$ v^{\rm ext}({\bf r}) \to n({\bf r})$] is well known to exist, because
$v^{\rm ext}({\bf r})$ determines the many-particle Hamiltonian,
Eq.~(\ref{defrel}) is less obvious.
In other words, Hohenberg and Kohn realized that for the ground state
the functional $n({\bf r}) = n[ \Psi] =
\langle \Psi | \sum_i \delta({\bf r} -
{\bf r}_i)| \Psi \rangle$
can be inverted, i.e., $\Psi = \Psi[n({\bf r})]$. 
With the help of this theorem the variational problem of the many-particle
Schr\"odinger equation transforms to a variational problem of an energy functional:
\begin{equation}
E_0 \leq \langle \Psi|H|\Psi \rangle = E_v[\Psi[n]] = E_v[n]\quad .
\label{variat}
\end{equation}
Here $E_0$ is the energy of the ground state, and
$E_v[n] = \int d{\bf r}\, v^{\rm ext}({\bf r}) n({\bf r}) + G[n]$.
In this functional $n({\bf r})$ is the variable (the electron ground-state
density of any $N$-electron system), and 
$v^{\rm ext}({\bf r})$ is kept fixed.
$G[n]$ is a {\em universal} functional independent of the
system, i.e., independent of $v^{\rm ext}({\bf r})$.
For example,  
$G[n]$ is the same for an H-atom, a CO-molecule, a solid etc.
The main advantage of this approach is that
$n({\bf r})$ only depends on three variables
while $\Psi(\{{\bf r}_i\})$
depends on many variables (the 3$N$ coordinates of all electrons).\refnote{13}
Thus, it is plausible that the variational problem of $E_v[n]$ is
easier to solve than that of $\langle \Psi|H|\Psi \rangle$, yet the result
for the ground-state energy and the ground state electron density
will be the same. The total energy entering Eq.~(\ref{PES}) is\refnote{14}
\begin{equation}
E^{\rm tot} ( \{{\bf R}_J \} ) = E_0( \{ {\bf R}_J \} ) +
\frac{1}{2}\sum_{J,J',J \ne J'} \frac{Z_J Z_{J'}}{|{\bf R}_J -
{\bf R}_{J'}|} \quad,
\label{tot}
\end{equation}
where $\{ {\bf R}_J\}$ includes all atoms, and $Z_J$ is the
nuclear charge.

\begin{figure}[tb]
\begin{center}
   \begin{picture}(200,200)
      \includegraphics{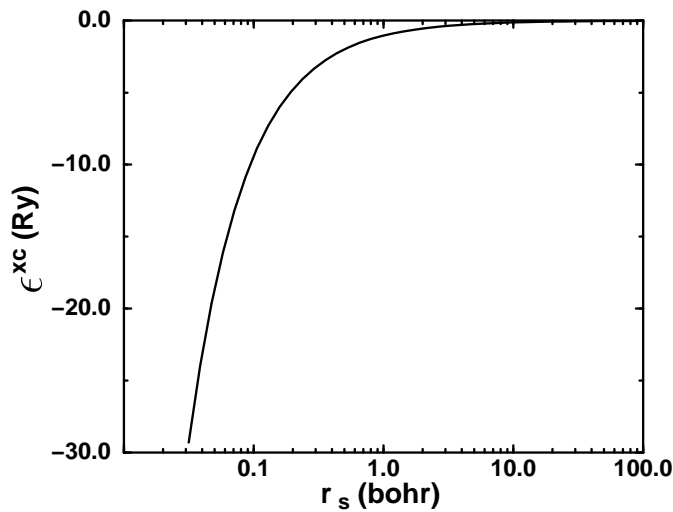}
   \end{picture}
\end{center}
\caption{Exchange-correlation energy per particle, $\epsilon^{\rm xc}$,
of homogeneous electron gases with density parameters $r_{\rm s}$. The electron
density and the density parameter are related by $n = \frac{4}{3} \pi r_s^3$.}
\label{excpar}
\end{figure}

An important problem remains, namely that
an explicit form of the functional $G[n]$ is unknown. Earlier work (in particular 
the Thomas-Fermi approach) had shown
that the treatment of the kinetic energy 
$\langle \Psi | -\frac{1}{2}\nabla^2 | \Psi \rangle$ is of particular 
importance and Kohn and Sham\refnote{15} therefore wrote the energy functional in the form
\begin{equation} 
E_v[n]  =  T_s[n] + \int d{\bf r}\, v^{\rm ext}({\bf r}) n({\bf r}) + 
\frac{1}{2}  \int d{\bf r}\,  v^{\rm H}({\bf r}) n({\bf r}) + E^{\rm xc}[n] \quad ,
\label{excdef}
\end{equation}
where $T_s[n]$ is the functional of the kinetic energy of a system of 
non-interacting electrons with density $n({\bf r})$, and
$v^{\rm H}({\bf r}) = \int d{\bf r'}\, \frac{n({\bf r'})}{|{\bf r}
 - {\bf r'}|}$ 
is the Hartree potential that describes the electrostatic 
interaction between electrons.
$E^{\rm xc}[n]$ is the so-called exchange-correlation
functional. It accounts for the Pauli principle, 
dynamical correlations due to the Coulomb repulsion,
and the correction of the self-interaction included 
for convenience in the Hartree term. With Eq.~(\ref{excdef}) the problem of the unknown
functional $G[n]$ is transformed to one 
involving $T_s[n]$ and $E^{\rm xc}[n]$. 
We note in passing that the functional defined by Eq.~(\ref{excdef}) can be
also modified by adding terms which vanish at the correct electron
density. Such a functional $E_v[n]$ may converge faster towards the
ground state or may depend less sensitive on the input density. The latter
implies that the input density does not need to be very good, yet the
resulting energy represents an acceptable approximation for the correct total
energy (see e.g., Ref.~16). Although the functional
$T_s[n]$ is not known explicitly in a mathematically closed form, it can 
be evaluated exactly by using the following ``detour'' proposed by Kohn and 
Sham. The variational principle applied to Eq.~(\ref{excdef}) leads to
\begin{equation}
\frac{\delta E_v[n]}{\delta n({\bf r})}  = 
\frac{\delta T_s[n]}{\delta n({\bf r})} + v^{\rm ext}({\bf r}) + v^{\rm H}({\bf r}) +
\frac{\delta E^{\rm xc}[n]}{\delta n({\bf r}) }
\nonumber 
\end{equation}
\begin{equation}
\hspace{-0.77cm}
 =  \frac{\delta T_s[n]}{\delta n({\bf r})} + v^{\rm eff}({\bf r}) = \mu \quad ,
\label{variational2}
\end{equation}
where $\mu$ is the Lagrange multiplier associated with the requirement 
of a constant particle number and thus equals the electron chemical
potential. The effective potential is defined as 
\begin{equation}
v^{\rm eff}({\bf r}) = v^{\rm ext}({\bf r}) + v^{\rm H}({\bf r}) +
v^{\rm xc}({\bf r}) \quad,
\label{eff_pot}
\end{equation}
with $v^{\rm xc}({\bf r}) = \delta E^{\rm xc}[n] / \delta n({\bf r})$,
and $n({\bf r})$ is a ground-state density of any  non-interacting electron system,
i.e., 
\begin{equation}
n({\bf r}) = \sum_{i=1} f_i | \phi_i({\bf r}) |^2 \quad,
\label{denans}
\end{equation}
where we introduced the occupation numbers $f_i$. Because
$T_s[n]$ is the kinetic energy functional of non-interacting
electrons, Eq.~(\ref{variational2}) (together with Eq.~(\ref{denans}))
is solved by
\begin{equation}
\left[- \frac{1}{2} \nabla^2 + v^{\rm eff}({\bf r}) \right] \phi_i({\bf r}) = 
\epsilon_i \phi_i({\bf r}) \quad.
\label{kseq}
\end{equation}
These are the Kohn-Sham equations,
that are to be solved self-consistently together with Eqs.~(\ref{eff_pot}) and
(\ref{denans}).
In principle this gives the exact ground-state electron
density and total energy of a system of interacting electrons.
However, the functional $E^{\rm xc}[n]$ is
still unknown. Some general properties of this functional and
values for some special cases are known.
Detailed and very accurate understanding exists for systems of constant
electron density. The asymptotic behavior at low and high
densities is given by expressions derived by Wigner\refnote{17}
and Gell-Mann and Brueckner\refnote{18} and for  intermediate
densities quantum Monte Carlo calculations have been performed
by Ceperley and Alder.\refnote{19} This gives the simple curve
shown in Fig.~\ref{excpar}, and this result for
$\epsilon^{\rm xc}(r_s) := \epsilon^{\rm xc}_{\rm LDA}(n)$ is then used 
in the functional
\begin{equation}
E^{\rm xc}_{\rm LDA} [n] = \int d{\bf r} \, n({\bf r}) 
\epsilon^{\rm xc}_{\rm LDA} (n({\bf r}))\quad ,
\label{lda.exc}
\end{equation}
which is the local-density approximation (LDA).\refnote{15}
Thus, in the LDA the many-body effects are included such that for
a homogeneous electron gas the treatment is exact and for
an inhomogeneous system exchange and correlation are treated by assuming that the system
can be composed from many small systems with a locally constant density.

The LDA  can be 
improved by including the dependence on the density
gradient which leads to the generalized gradient approximation (GGA).
Several different GGA's were proposed in the literature
\refnote{20-25} and have been used successfully 
for DFT calculations of atoms, molecules, bulk solids, and surfaces (an 
overview can be found in Refs.~25 and 26), but also limitations 
have been identified for example by Mitas {\it et al.}\refnote{27}
and Umrigar and coworkers.\refnote{28}
It is by now clear that the lattice constants calculated with a
GGA are typically larger than those obtained with the LDA,
with the experimental values  usually being in between. 
Binding energies (or cohesive energies) of
molecules and solids are clearly improved by the GGA and
energy barriers of chemical reactions are improved as well
(see Ref.~29 and 30 and references therein).
Still, for surface diffusion DFT-LDA calculations give
energy barriers in good agreement with those deduced from experiments
and with GGA calculations. Although the total energies are changed
when going from the LDA to the GGA,
the changes in energy barriers, i.e., in total energy {\em differences},
are typically less pronounced (see e.g., Ref.~31).\\

\section{IMPLEMENTATION OF DENSITY-FUNCTIONAL THEORY TO COMPUTE 
MICROSCOPIC PARAMETERS}

\begin{figure}[b]
\leavevmode
\includegraphics{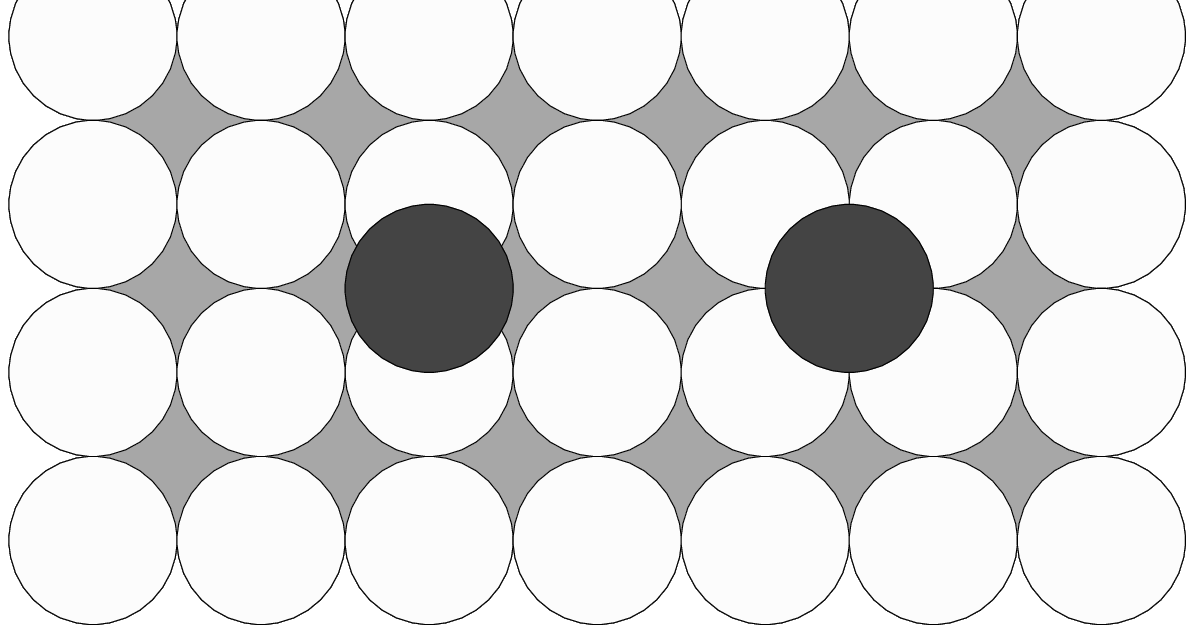}
\includegraphics{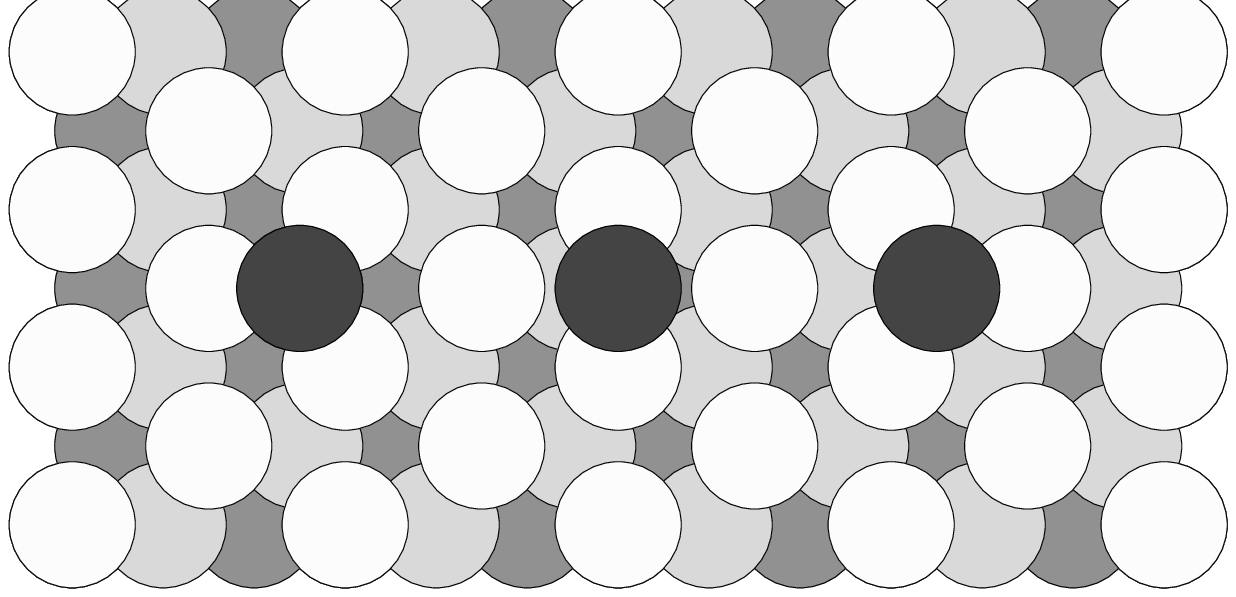}
\put(182,-133){\vector(0,1){60}}
\put(286,-133){\vector(0,1){60}}
\put(234,-143){\vector(0,1){70}}
\put(213,-295){\vector(0,1){50}}
\put(267,-310){\vector(0,1){65}}
\put(165,-145){hcp-site}
\put(270,-145){fcc-site}
\put(210,-155){bridge-site}
\put(185,-307){bridge-site}
\put(240,-332){hollow-site}
\put(100,-40){\bf (111)}
\put(110,-220){\bf (100)}
\vspace*{1cm}
\caption{
Top view at a fcc\,(111) and (100) surface. The adsorption sites 
labeled fcc, hcp, and hollow site usually
correspond to the most stable binding sites 
while the bridge-site is the transition state of a hopping diffusion.}
\label{ad_sites}
\end{figure}

Typically there are only a few candidates for 
adsorption sites and possible channels for diffusion. This is illustrated in 
Fig. \ref{ad_sites}
for the fcc\,(111) and fcc\,(100) surfaces.
For adatoms which are chemically similar (or equivalent) to those of the
substrate the stable sites are those with high coordination and
for hopping diffusion the transition state
is at the bridge site. The relevant information about the PES then
is obtained by calculating the total 
energy of the system with the adatom placed in those positions. 
In general, more care is necessary because the bridge site can also be a local 
minimum of the PES and the energy barrier could be in between the 
high coordination and the bridge sites. Furthermore, it is possible
that the diffusion does not proceed by hopping but by atomic 
exchange.\refnote{3-7}

\begin{figure}[tb]
\leavevmode
\includegraphics{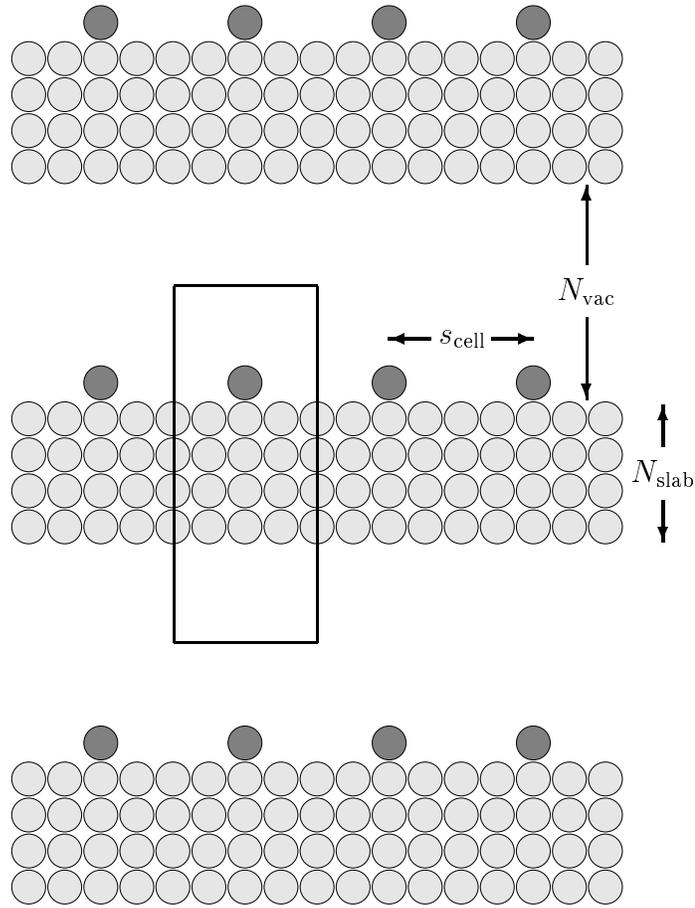}
\vspace*{13cm}
\caption{
Sketch of a supercell describing  an ``isolated'' adatom
at a surface (side view).}
\label{supercell}
\end{figure}

In the bulk crystal the three-dimensional periodicity can be
exploited by using Bloch's theorem. Unfortunately,
the presence of a surface and an adatom on top of it breaks all
periodicity. The (in principle) best approach to treat such
difficult situation is given by the Green-function
method.\refnote{32,33}
A popular approximation for (at least in the past)
for adsorbate systems is the cluster approach.\refnote{34}
The presently most efficient and practical
approach that was also used in the results discussed below
is the supercell approach. The supercell may be also called
a big cluster, but in contrast to conventional cluster calculations
the supercell is periodically repeated.
As a consequence, the cluster boundary is treated physically
very accurately, and by utilizing the
periodicity, i.e., the Bloch theorem, it is possible to use
very big cells.
The idea of an adatom on top of a substrate in the supercell approach is 
sketched in Fig. \ref{supercell}.
The adatom is placed on top of a slab of a certain number of layers.
The number of layers $N_{\rm slab}$ 
must be sufficiently large 
so the adatom does not 
``feel'' the presence of another surface on the other side of the slab
(or at least that the quantity to be computed, e.g., a diffusion barrier,
is not affected by the other surface).
Alternatively,
one could also place an adatom on either side of the slab; in this case,
there are more symmetries in the geometry but more layers are needed in the
slab to screen the mutual interaction between the two adatoms
through the slab.
The adequate number of layers $N_{\rm slab}$ depends on the properties 
that one wants to calculate 
and the surface orientation,
and careful tests 
must be carried out. 
For example, for the Ag\,(111) surface 
four layers are sufficient when the adatom is placed on only one side 
of the slab, while for Al\,(100) seven layers are necessary.

As illustrated in Fig. \ref{supercell} the geometry repeats periodically in
vertical and lateral directions. The lateral periodicity implies that a single
adatom placed on a substrate is not at all a single adatom; if the cell size
parallel to the surface is chosen, for example, as ($2\times2$) we actually
calculate a system with a coverage of $25\,\%$. It is therefore important to
test that the interaction with the neighboring adatoms can be neglected. On a
(111) surface a cell size of ($2\times2$) is usually sufficient, but sometimes
larger cells [($3\times3$) or even ($4\times4$)] are necessary.  To model a
diffusion event along or across a step one either chooses a small island on
top of a substrate or a vicinal surface. The latter has the advantage that
only one step exists in the unit cell so a smaller cell size is required to
attain a negligible step-step interaction. The system also repeats in the
vertical direction separated by a vacuum region.  The thickness of the vacuum
region must be tested as well, but the computational cost of a thicker vacuum
region is relatively small compared to a larger cell size or a higher number
of slab layers (for a deeper discussion of the above technicalities, see e.g.,
Ref.~35).

Core electrons typically do not take part directly
in the binding process of atoms in molecules
and solids, and the nature of the chemical bond is mainly determined by 
the valence electrons. This is exploited by the {\it frozen core
approximation} where the core electrons are effectively combined with the 
nuclei to form frozen (i.e. unpolarizable) ions.
Still, not just the electrostatic potential but also the quantum nature of the core
electrons is felt by the valence electrons. For example, different wave
functions must be orthogonal and therefore the valence wave functions have
nodes and oscillate in the core region.
For practical calculations one needs to expand the wave function
in a suitable basis
and we choose a plane wave basis set\refnote{36}
\begin{equation}
|\phi_j({\bf k,r})\rangle =  \sum_{{\bf G}}
c_{j,{\bf k}}({\bf G}) |{\bf G+k} \rangle \quad .
\label{plane_wave}
\end{equation}
A plane wave description of wave functions that have nodes and oscillate requires a
very large number of plane waves. This inconvenience is cured efficiently by
the {\it pseudopotential} approach.
Modern {\em ab initio} pseudopotentials reproduce the
potential of an atom exactly outside the core region defined by a radius $r_c$ 
and are rather smooth
inside the core region.
An important requirement on a ``good'' pseudopotential
is that it is transferable, which means that the pseudopotential
should behave like the all-electron potential in a variety of
different chemical situations. 
Pseudopotentials that reproduce the 
same charge inside the core region as the all-electron  potential,
and therefore have the same scattering properties, 
are referred to as {\it norm-conserving}.
Those that are often used have been developed by
Bachelet, Hamann, and Schl\"{u}ter\refnote{37}, Troullier and
Martins\refnote{38}, and Gonze, Stumpf, and Scheffler.\refnote{39}
Recently, Vanderbilt \refnote{40}
proposed {\em ab initio} pseudopotentials that 
drop the condition of norm-conservation and therefore can be used with a lower
number of plane waves. The gain in computer time due to the smaller basis set
is partially compensated by the costs to calculate the correction required by
the neglected norm-conservation.

The electron density is calculated according to Eq.~(\ref{denans}) as
\begin{equation}
n({\bf r}) = \sum_{\bf k} \sum_j \omega_{\bf k} f(\epsilon_j({\bf k})) |\phi_j({\bf k,r})|^2
\label{ksum}
\end{equation}
where the integration over the Brillouin zone has been replaced by a sum over
a mesh of ${\bf k}$-points with $\omega_{\bf k}$ the ${\bf k}$-point's weight.
Such replacement typically works quite efficiently because the electron wave
functions vary rather smoothly with ${\bf k}$, and the main dependence usually is
in the phase-factor. Thus, a certain region in ${\bf k}$-space
is well represented by only one ${\bf k}$-point.
A convenient scheme to construct an appropriate {\bf k}-point mesh
is described by Monkhorst and Pack.\refnote{41}
In {\em ab initio} pseudopotential calculations  some matrix elements and
some integrals are efficiently evaluated in real space whereas
others are efficiently evaluated in
reciprocal space. The technique of fast Fourier transformation
enables a numerically fast change from one representation  to the other.
Technical details of the computational procedures
are described for example in Refs.~8 and 10.

\section{RESULTS FOR MICROSCOPIC PARAMETERS}

We now discuss some results
obtained by DFT calculations that provide a deeper insight into
the microscopic mechanisms behind growth phenomena.
The main objective is
to identify the nature and to determine the energetics of diffusion
processes. 
For the homoepitaxial growth of metallic systems such as Al/Al\,(111),\refnote{35,42}
Al/Al\,(100),\refnote{35,43} and Ag/Ag\,(100)\refnote{7,31} 
comprehensive DFT studies have determined diffusion barriers
for diffusion on the flat surface and along and across step edges with 
the hopping and the exchange mechanism. In this section the properties of
Ag/Ag\,(111) are discussed and the next section extends earlier DFT calculations
of Al/Al\,(111)\refnote{35,42} and analyzes actual growth conditions of
mesoscopic islands on long time scales.

\begin{figure}[b]
\unitlength1cm
\begin{center}
   \begin{picture}(7,6)
      \includegraphics{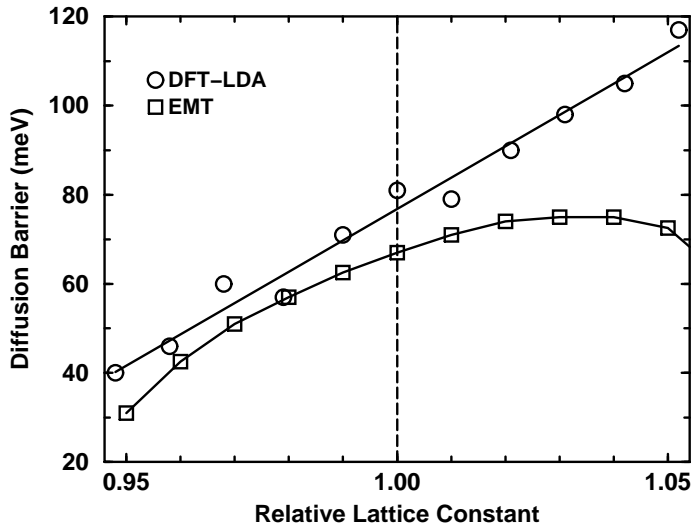}
   \end{picture}
\end{center}
\caption{
Diffusion barrier (in meV) for Ag on Ag\,(111) as function of strain.
The circles are DFT-LDA results from Ref.~47
and the squares are EMT results from Ref.~44.}
\label{Strain_barrier}
\end{figure}
Growth of one material on a different material is of particular interest for
a number of technological applications. In such a heteroepitaxial system
with usually different lattice constants
the material to be deposited is under the influence
of epitaxial strain.
Growth of Ag on Pt\,(111) and Ag on a thin Ag film on Pt\,(111) has been
the focus of a number of recent studies,\refnote{44-46}
and with a lattice mismatch of $4.2\,\%$ it 
provides important information 
about the effects of strain during growth.
Here, we will particularly discuss how strain affects the surface 
diffusion barrier.

Only few theoretical studies of the influence of
lattice mismatch on the diffusion barrier are present in the literature. 
\begin{table}[b]
\centering
\begin{tabular}{|l|c|c|c|}
\hline\hline
System & Experiment (Ref.~44) & EMT (Ref.~44) &
DFT (Ref.~47) \\
\hline
Ag/Pt\,(111) & 157 & 81 & 150 \\
Ag/1ML Ag/Pt\,(111) & 60 & 50 & 65 \\
Ag/Ag\,(111) & 97 & 67 & 81 \\
\hline
\end{tabular}
\caption{Diffusion barriers (in meV) for Ag on Pt\,(111), Ag on 
one monolayer (ML) Ag on Pt\,(111), and Ag on Ag\,(111).}
\label{Ag_Pt_barriers}
\end{table}
For a metallic system we are only aware of results for Ag on Ag\,(111) where
the authors of  Ref.~44 find in a semi-empirical 
effective medium theory (EMT) calculation that the diffusion barrier increases under
tensile strain and decreases under compressive strain.

Here, we present first-principles calculations (more details are given in
Ref.~47) studying the dependence of the diffusion barrier
on the lattice constant for Ag on Ag\,(111).\refnote{44} In the
range of $\pm 5\,\%$ strain the DFT results exhibit an approximately linear
dependence with a slope of $\sim 0.7$ eV (see Fig. \ref{Strain_barrier}).  The
calculated diffusion barrier for the unstrained system, $E_{\rm
  d}^{\rm{\mbox{\scriptsize{Ag-Ag}}}} = 81$~meV, is in good agreement (within
the error margins of the experiment and the calculations) with the scanning
tunneling microscopy (STM) results of $E_{\rm
  d}^{\rm{\mbox{\scriptsize{Ag-Ag}}}} = 97$~meV. The accordance between
experiment and theory extends to the system Ag/Pt\,(111) and Ag/1ML
Ag/Pt\,(111). These results are summarized in \mbox{Table
  \ref{Ag_Pt_barriers}.} In Fig. \ref{Strain_barrier} the DFT-LDA results are
compared to those of an EMT study.\refnote{44} The EMT results
exhibit a linear dependence only for very small values of strain ($\pm2\,\%$)
and the diffusion barrier starts to decrease for values of misfit larger than
$3\,\%$.  Indeed, it is plausible that a decrease of the diffusion barrier
occurs when the atoms are separated far enough that eventually bonds are
broken.  However, as our DFT-LDA results show, for Ag/Ag\,(111) this happens
at values for the misfit that are larger than $5\,\%$.  Additionally, when
comparison with experiment is possible [i.e., Ag on Ag\,(111), and Ag on a
monolayer Ag on Pt\,(111)] the EMT results are off by a factor that varies
from 1.2 to 2.

The DFT results in Fig. \ref{Strain_barrier} were obtained with the LDA for
the exchange-correlation functional and test calculations show that GGA
increases the diffusion barrier by no more than $5 - 10 \,\%$.  This has also
been found for Pt on Pt\,(111) and Ag on Ag\,(100)
(Refs.~48 and 31, respectively).  The
general trend of an increasing energy barrier for hopping diffusion with
increasing lattice constant is quite plausible (for exchange diffusion, see
Ref.~7).  Smaller lattice constants correspond to a
reduced corrugation of the surface, and as a result of a large compression the
atom is not bonded much stronger at the adsorption sites than at the bridge
site.  In contrast, when the surface is stretched the corrugation increases
and the adsorption energy at the three-fold coordinated hollow sites
increases. This picture will change when the strain is so large that bonds are
broken and then it is expected that the hopping diffusion barrier will start
to decrease again at very large tensile strain.

It is worth noting that the diffusion barrier for Ag on top of a
pseudomorphic layer of Ag on Pt\,(111) is substantially lower
than that for Ag on Ag\,(111). A question that arises is whether 
this reduced diffusion barrier is a result of the compressive strain or 
electronic effects due to the Pt substrate.
The diffusion barrier for Ag on Ag\,(111) with a lattice constant that is compressed to 
the value of the lattice constant for Pt is 
$E_{\rm d}^{\rm{\mbox{\scriptsize{Ag-Ag}}}} = 60$~meV while that for 
Ag on Pt\,(111) (also with the Pt lattice constant 
of 3.92 \AA \, obtained from DFT) 
is $E_{\rm d}^{\rm{\mbox{\scriptsize{Ag-Ag/Pt}}}} = 65$~meV. The agreement of these two values
suggests that the reduction of the diffusion barrier for
Ag on a layer of Ag on Pt\,(111) is mainly a strain effect and that the diffusion
barrier on top of a layer of Ag is essentially independent of the substrate
underneath. 

We note in passing that the increase of the hopping diffusion barrier with
tensile strain is also to be expected (and found) for the more open (100)
surface. On the other hand, for exchange diffusion the slope of the energy
barrier as a function of strain has the opposite sign and it has been
argued\refnote{7} that this behavior and the large surface stress at
late $5d$ transition metals actuate exchange diffusion experimentally found
for Ir\,(100) and Pt\,(100).

\section{{\em{AB INITIO}} KINETIC MONTE CARLO SIMULATIONS}

The time between two successful diffusion events is often of the order of
nanoseconds. Since molecular dynamics (MD) calculates all
unsuccessful attempts (usually $\sim 10^{3}$) explicitly, a typical
MD simulation can cover at most times of some picoseconds, possibly some
nanoseconds.  
Therefore, although MD simulations can provide important insight into 
elementary microscopic mechanisms, they normaly cannot be used for growth studies.
Instead, the method of choice for studying the spatial and temporal development of
growth is kinetic Monte Carlo (KMC).
The key idea behind KMC is to describe stochastic processes (such as
deposition, diffusion,
desorption, etc.) on the microscopic
scales by rates and thus to avoid the explicit calculation of
unsuccessful attempts. Yet, the result of a KMC study will be
the same as that of an MD simulation, provided that the underlying PES is the same.  
The strategy of KMC can be summarized as follows:
\begin{enumerate}
\item[$1)$] Determination of all processes $j$ that possibly could take place with
  the actual configuration of the system.
\item[$2)$] Calculation of the total rate $R = \sum_j \Gamma^{(j)}$, where the sum
  runs over the possible processes [see step 1)]. Deposition
  is accounted for in this description by the deposition rate $F\quad .$
\item[$3)$] Choose a random number $\rho$ in the range $(0,1]$.
\item[$4)$] Find the integer number $l$ for which
\begin{equation}
\sum_{j = 1}^{l - 1} \Gamma^{(j)} \leq \rho \, R < \sum_{j = 1}^{l}
\Gamma^{(j)}\quad .
\label{mcalgo}
\end{equation}
\item[$5)$] Let process $l$ to be taken place.
\item[$6)$] Update the simulation time $t := t + \Delta t$ with 
$\Delta t= -ln(\rho)/R\quad .$
\item[$7)$] Go back to step $1)$.
\end{enumerate}

KMC simulations have been used to study crystal growth of 
semiconductors (e.g., Refs.~49-51)
and metals (e.g., Refs.~52-55).
However, most of these studies have been based on restrictive approximations. 
For example, the input parameters, such as activation barriers, have been
treated as effective parameters determined rather indirectly, e.g. from the fitting
of experimental quantities, like intensity oscillations in helium atom
scattering (HAS) measurements, in reflection high energy electron
diffraction (RHEED) experiments, or island densities from STM pictures. Thus,
the connection between these parameters and the microscopic nature of the
processes may be somewhat uncertain. 
Often even the surface
structure was treated incorrectly, i.e., the simulation was done on a simple
cubic lattice while the system of interest had an fcc or bcc structure.
Despite these approximations, 
such studies have provided
significant qualitative and in some cases also quantitative
insight into growth phenomena. The next better approach is to use 
semi-empirical calculations such as the 
embedded atom method
or effective medium theory to evaluate the PES 
for KMC simulations of growth.\refnote{56-58}
The best, but also most elaborate approach to obtain the PES was 
described in the previous sections. 
In the following  the DFT results for Al on Al\,(111) obtained by Stumpf 
and Scheffler\refnote{35,42} are utilized for KMC simulations.
Thus, it is our aim to perform a realistic simulation that takes 
into account the correct structure of the system and  
rate constants determined from accurate {\it ab initio} calculations.

\begin{figure}[tb]
\leavevmode
\includegraphics{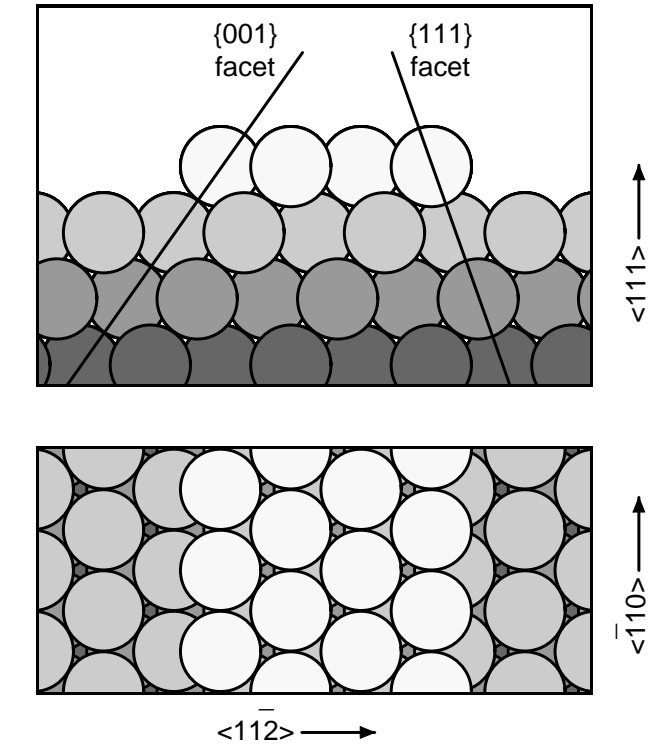}
\vspace*{10cm}
\caption{
Side view (upper panel) and top view (lower panel) of the two types of
close-packed steps on the \,(111) surface
of an fcc crystal.}
\label{steps111}
\end{figure}

On the \,(111) surface of an fcc crystal there are two
different types of close-packed steps, shown in Fig.~\ref{steps111}.
They are labeled according to their $\{100\}$ and $\{111\}$ facets, referring
to the plane passing through the adatom of the step and the neighbor atom of the 
substrate (often these two steps are labeled A and B, 
respectively). 
Experimentally it has been shown that for Pt\,(111)\refnote{59} and
Ir\,(111)\refnote{60} these two steps behave quite differently with respect to
surface diffusion and growth. 
For Al\,(111) the DFT calculations\refnote{35,42} predict that the formation energies of the two steps
are different with a lower energy cost for
the formation of the $\{111\}$ faceted  step than of the $\{100\}$ faceted step: 0.232 eV per atom
vs. 0.248 eV per atom. This difference affects the equilibrium shape of the
islands as determined by the Wulff construction. Because more open
steps have a higher formation energy, one expects in thermodynamic equilibrium
and at not too high temperatures 
hexagonally shaped islands where the edges alternate between those with a $\{100\}$ and a
$\{111\}$ microfacet, the latter being longer. 

We now like to analyze typical growth conditions, i.e., a situation
far from equilibrium where kinetic processes
are dominant. For Al on Al\,(111) Stumpf and Scheffler\refnote{35,42} analyzed
microscopic diffusion processes and in particular determined 
the activation energies $E_{\rm d}$ for: 
\begin{enumerate}
\item[$(i)$] diffusion of a single adatom on the flat surface: $E_{\rm d}$ = 0.04 eV;
\item[$(ii)$] diffusion from upper to lower terraces 
which was found to proceed by an 
exchange with a step-edge atom:
  $E_{\rm d}$ = 0.06 eV for the \{100\} step and $E_{\rm d}$ = 0.08 eV for the \{111\} step;
\item[$(iii)$] diffusion parallel to the \{100\} step via hopping: $E_{\rm d}$ = 0.32 eV
  (0.44 eV for exchange);
\item[$(iv)$] diffusion parallel to the \{111\} step via exchange: $E_{\rm d}$ = 0.42 eV
  (0.48 eV for hopping).
\end{enumerate}

\begin{figure}[tb]
  \leavevmode
  \includegraphics{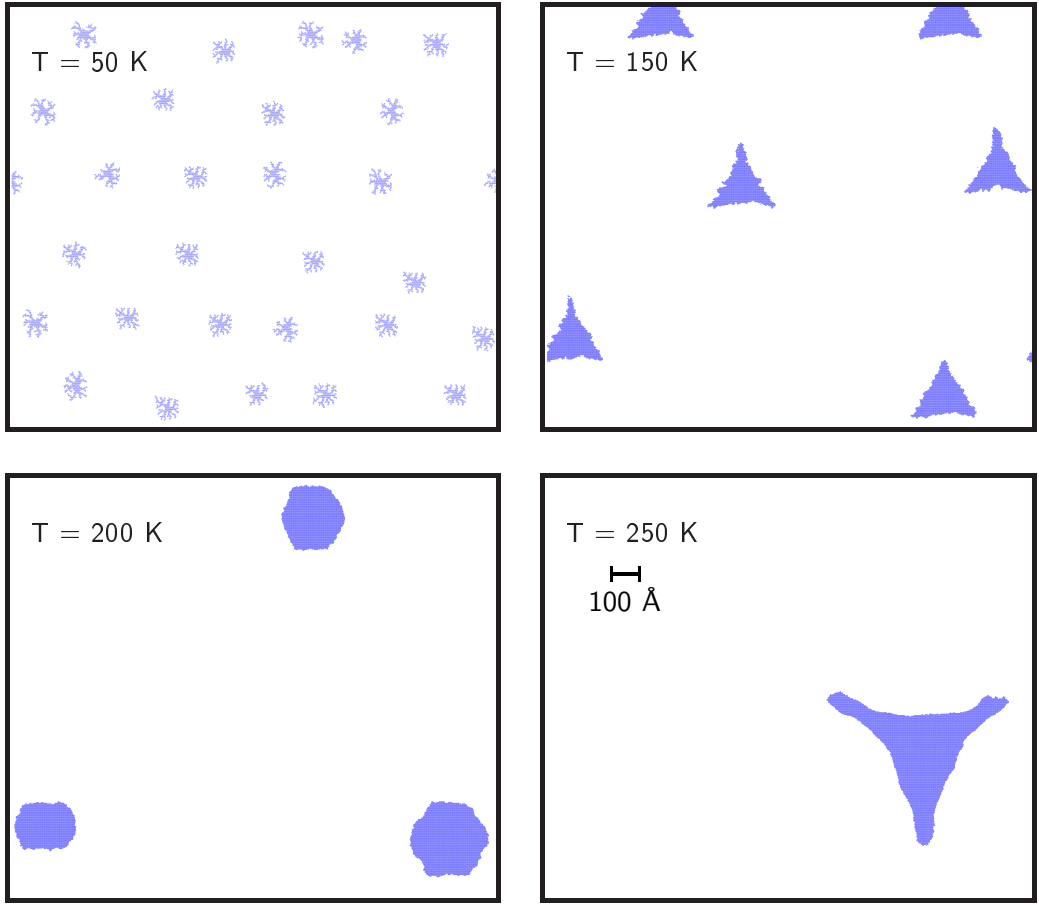}
\unitlength1cm
\begin{picture}(0,0)
\thicklines
\put(9,-8){\line(1,0){0.38}}
\put(9,-8.1){\line(0,1){0.2}}
\put(9.38,-8.1){\line(0,1){0.2}}
\put(8.7,-8.5){\small{\textsf{100 \AA}}}
\end{picture}
\vspace*{13.0cm}
\caption{A surface area of (1718 $\times$ 1488)~\AA$^2$~ (half of the
  simulation array) at four different substrate temperatures.
The deposition rate was 0.08 ML/s and the coverage in each picture is $\Theta$ = 0.04 ML.}
\label{fig.2}
\end{figure}

The DFT calculations\refnote{35} give that the binding energy 
of two adatoms in a dimer is 0.58 eV, and we therefore assume that 
dimers, once they are formed, are stable, i.e., they will not dissociate. 
Moreover, in the lack of reliable information we assume that dimers are immobile.
We note that the reported value\refnote{35,42} for the self-diffusion energy barrier is 
rather low (0.04 eV) and comparable to the 
energy of optical phonons
of Al\,(111) ($0.03 - 0.04$ eV).\refnote{61} Thus, simulations at room
temperature may not be 
reliable because the concept of single jumps between nearest 
neighbor sites
is no more valid. A single optical phonon can furnish enough energy to 
an adatom for leaving its adsorption site and diffusing on the flat surface. At room temperature
the level population of optical phonon is high and 
the adatoms have practically no saddle
point and migrate freely on the flat surface. We therefore limited our study to
substrate temperatures  $T \leq 250$ K. 

We adopt periodic boundary conditions, and our rectangular simulation area is
compatible with the geometry of an fcc\,(111) surface. The dimensions of the
simulation area are 1718 $\times$ 2976~\AA$^2$. These dimensions are a
critical parameter and it is important to ensure that the simulation area is
large enough that artificial correlations of neighboring cells do not affect
the formation of growth patterns.  The mean free path $\lambda$ of a diffusing
adatom before it meets another adatom with possible formation of a nucleation
center or is captured by existing islands should be smaller than the linear
dimension of the simulation array. Since $\lambda$ is proportional to
$(D/F)^{1/6}$ (Ref.~62), we have that (with $F = 0.08$
ML/s) $\lambda \sim 50$ \AA~ for $T$ = 50 and gets as large as $\sim 10^3$
\AA~ for $T$ = 250 K. We see that our cell is large enough (for the imposed
deposition rate) for $T \leq 150 K$, whereas at higher temperatures the
dimensions of the cell are too small, i.e., for $T > 150 K$ the island density
is determined by the simulation array rather than the physics. Nevertheless,
the island shape is determined by local processes (edge diffusions) and is
still meaningful.

In the KMC program two additional insights extracted from
the DFT calculations are included: $(i)$ the attractive interaction between steps
and single adatoms, and $(ii)$ the fact that diffusion processes take place via
different mechanisms (hopping or exchange). Particularly the second point
plays an important role in our investigation. In several KMC
simulations of
epitaxial growth the attempt frequency of the diffusion rate
has the same value
for all the processes, and this value lies usually in the range of a typical
optical phonon vibration or the Debye frequency. However, this 
assumption may not be right. First,
processes with larger activation barriers may have a 
larger attempt frequency than processes with smaller energy barriers. This is a
consequence of the compensation effect described, e.g., in
Ref.~63. Moreover, processes as hopping and exchange that involve a
different number of particles and different bonding configurations may also be
characterized by different
attempt frequencies. This has been observed~\refnote{64-67}
for several systems (Rh, Ir, Pt) and implies 
that the attempt frequency for exchange
diffusion can be larger by up to two orders of magnitude
than that for hopping.
\begin{figure}[b]
\unitlength1cm
\begin{center}
   \begin{picture}(7,6)
      \includegraphics{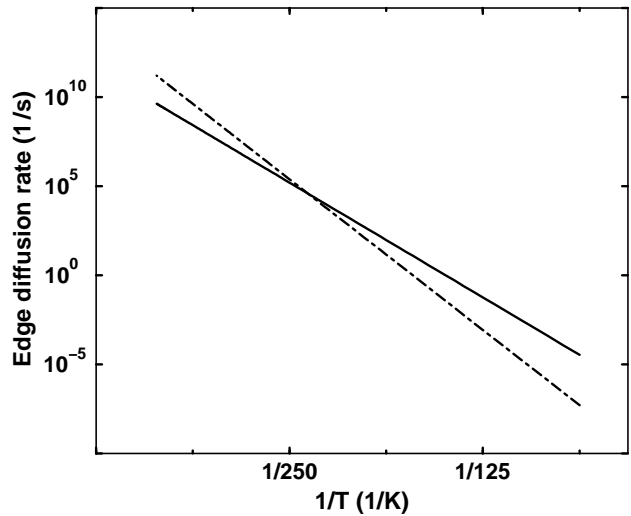}
   \end{picture}
\end{center}
\caption{Temperature dependence of the edge diffusion rates for atom
  diffusion along the \{100\} step by hopping with $\Gamma_0$ = $2.5 \times
  10^{12}$ s$^{-1}$
  (solid line), and along the \{111\} step by exchange with $\Gamma_0$ = $2.5 \times 10^{14}$ s$^{-1}$
  (dash-dotted line).}
\label{fig4}
\end{figure}
For Al surfaces calculations with the embedded atom method\refnote{68} showed
a difference of prefactors of one order of magnitude. 

The results of the {\it ab initio} KMC simulations are shown in
Fig.~\ref{fig.2} for coverages of $\Theta = 0.04$ ML.  When the substrate
temperature is 50~K during growth the shape of the islands is highly irregular
and indeed fractal. Adatoms which reach a step cannot leave it anymore and
they even cannot diffuse along the steps.  Thus, at this temperature
ramification takes place into random directions, and island formation can be
understood in terms of the so-called {\it hit and stick} model (see also
Ref.~69).  At a growth temperature $T$ = 150~K the island
  shapes are triangular with their sides being \{100\} steps. Increasing the
  temperature to $T$ = 200~K a transition from triangular to hexagonal shape
  occurs and for $T$ = 250~K the islands become triangular again.  However, at
  this temperature they are mainly bounded by \{111\} steps.

To understand the island shapes in the temperature regime between 150
and 250~K, we consider the mobility of the adatoms along the steps (at such
temperatures the adatoms at the step edges cannot leave the steps): 
The lower the migration probability along a given
step edge, the higher the step roughness and faster the speed of advancement 
of this step edge. As a consequence, this step edge shortens as a result
of the growth kinetics and eventually it may even disappear.
Since diffusion along the densely packed steps on
the (111) surface (the \{100\} and \{111\} facets) is faster than along 
steps with any other orientation this criterion explains the presence of
islands which are mainly bounded by \{100\} or \{111\} steps. The same
argument can be extended to the diffusion along the two close-packed steps and
applied to the triangular islands at $T$ = 150~K, where the energy
barrier for the diffusion along the \{111\} facet is larger 
and thus the \{100\} steps 
survive so that triangular islands with \{100\} sides are obtained. By 
considering the energy barriers we would expect only these islands,
until the temperature regime for the thermal equilibrium is reached.
However, as noted in the introduction, the diffusion of adatoms is not only governed by the energy
barrier but also by the effective attempt frequency.
For Al/Al\,(111) the effective attempt frequencies have not been calculated, but the 
analysis of Ref.~35 proposes that the exchange process should 
have a larger attempt frequency than the hopping process.
The results displayed in Fig. ~\ref{fig.2} are obtained with 
$1.0 \times 10^{12}$ s$^{-1}$ for the
diffusion on the flat surface, $2.5 \times 10^{12}$ s$^{-1}$ for the
jump along the \{100\} step, and $2.5 \times 10^{14}$ s$^{-1}$ for the
exchange along the \{111\} step. 
These effective attempt frequencies are the only input of the KMC not calculated explicitly
by DFT, but were estimated from the theoretical PES as well as from 
experimental data for other systems.
In  Fig.~\ref{fig4} the edge diffusion rates along the two steps are plotted as
a function of the reciprocal temperature.
At lower temperatures the energy barrier dominates the diffusion rate, but at
$T$ = 250 K the attempt frequencies start to play a role 
and lead to faster diffusion along the \{111\} facet than along 
the \{100\} one.
Thus, the latter steps disappear and only triangles with \{111\} sides are
present. The roughly hexagonally shaped islands at $T$ = 200 K are a
consequence of 
the equal advancement speed for the two steps at that temperature. 
Obviously, the temperature-dependence of the growth shapes 
found in Fig. ~\ref{fig.2} is crucially determined by the ratio of 
the two diffusivities and in particular by the 
temperature at which the two lines of Fig.~ \ref{fig4} cross. 
If the difference were
only one order of magnitude, 
the crossing would be at a temperature that is too high (namely at $T = 505$
K). 
The formations of fractals (Fig. ~\ref{fig.2}, upper left) and of \{100\} step triangles
would still occur.
Obviously, the importance of the attempt frequencies should receive a better assessment
through accurate calculations, and work in this direction is in
progress. 
While no experimental data for Al/Al\,(111) 
are presently available we note that a similar sequences of islands as obtained above has been
observed for Pt on Pt\,(111) by Michely {\it et al.}.\refnote{59}


\begin{numbibliography}
\bibitem{units} All equations are noted in Hartree atomic units, i.e.,
the unit of length is 1 Bohr = 0.5292 \AA, the unit of energy
is 1 Hartree = 27.2116 eV, and $\hbar = m = e = 1$.
\end{numbibliography}
\end{document}